# Predicting Hospital Re-admissions from Nursing Care Data of Hospitalized Patients

Muhammad K Lodhi[1], Rashid Ansari[1], Yingwei Yao[2], Gail M Keenan[2], Diana Wilkie[2], Ashfaq A Khokhar[3]

[1] University of Illinois at Chicago, Chicago, IL, USA
{mlodhi3, ransari}@uic.edu
[2] University of Florida, Gainesville, FL, USA
{yyao, gkeenan,diwilkie}@ufl.edu
[3] Illinois Institute of Technology, Chicago, IL, USA
ashfaq@iit.edu

**Abstract.** Readmission rates in the hospitals are increasingly being used as a benchmark to determine the quality of healthcare delivery to hospitalized patients. Around three-fourths of all hospital re-admissions can be avoided, saving billions of dollars. Many hospitals have now deployed electronic health record (EHR) systems that can be used to study issues that trigger readmission. However, most of the EHRs are high dimensional and sparsely populated, and analyzing such data sets is a Big Data challenge. The effect of some of the well-known dimension reduction techniques is minimized due to presence of non-linear variables. We use association mining as a dimension reduction method and the results are used to develop models, using data from an existing nursing EHR system, for predicting risk of re-admission to the hospitals. These models can help in determining effective treatments for patients to minimize the possibility of re-admission, bringing down the cost and increasing the quality of care provided to the patients. Results from the models show significantly accurate predictions of patient re-admission.

## 1 Introduction

Sparse and high dimensional datasets pose a serious challenge to existing data mining and machine learning methods, mainly because of their size and exponential complexity w.r.t dimensions. Due to these characteristics, the gap between our ability to process and analyze data and the rate at which it is accumulating is rapidly widening [1]. These data sets stem from diverse application areas, such as electronic health records (EHRs), biology, astronomy, web data, and medical imaging. Due to the presence of non-linear variables and their varying degree of importance in different domains, the problem is complex and extremely challenging. Different data mining techniques have been used to extract knowledge available in some of these data sets, albeit with limited success until now [2]. Various algorithms [3, 4] have been introduced, that use row-wise enumeration method instead of traditional column-wise enumeration method to address the dimensionality problem, however, they have their own limitations as they work best for dense high dimensional datasets, with significantly lower number of rows compared to number of columns. Different dimension reduction algorithms, such as Principal Component Analysis [5], Multi-Dimensional scaling [6] and Independent component analysis [7], are too restrictive due to their reliance on global linearity assumption. The context of various variables is also abated using these conventional dimensional reduction methods.

In this paper, we target the analysis of a high dimensional and sparse dataset that stems from nursing care EHRs. In such a dataset, thousands of variables consisting of vitals, drugs, tests, treatments, etc. exist (high dimensionality), yet only a limited number of them are tracked for any individual patient (sparse). While nursing care data are an important part of the EHR, it usually goes unnoticed when planning for the improvement of healthcare delivery systems [8]. Effective use of different data mining methods can be extremely helpful in provisioning of better care to the patient, and developing more effective care, and consequently help in decreasing the healthcare cost.

Historically, the data stored in EHRs has been used to monitor the progress of the patients, though recently, a lot of research is being performed to build predictive models using this data. We believe that EHR systems are also a perfect candidate to study big data issues due to size and heterogeneity of the data. Other industries have been using the big data methods to save costs. On the other hand, more than a trillion dollars are wasted annually in healthcare industry partly due to latest technology not being used to its fullest in healthcare [9]. Big data techniques can have a huge impact in reducing the healthcare costs that are expected to continue to markedly increase in the coming years [10].

One of the reasons of the increasing costs of healthcare are the patient re-admissions to the hospitals. Reducing repeat hospitalizations can greatly reduce these costs. As with many other issues in the healthcare system, repeat hospitalizations often occur due to poor treatment provided to the patients [11, 12], more specifically, they are often caused by premature discharges [13] or communication breakdown between the patients and healthcare team while the patient is being discharged. These readmissions result in higher costs to taxpayers [14], costing as much as $45 billion annually [15, 16]. Medicare, along with other healthcare payers, are concerned with the cost of

unnecessary readmissions as Medicare alone spends roughly $15 billion annually on repeat hospitalizations [17] and almost 20% of the patients are readmitted within 30 days after being discharged from the hospitals [18]. According to a report, almost 76% of repeat admissions can be avoided by improving care before and after the patient is discharged [19]. By decreasing these preventable repeat hospitalizations, overall productivity of the hospitals and staff can improve considerably [20, 21].

Avoidable re-admissions are a huge burden on hospital resources, including the workforce. In this study, our aim is to determine different nursing and patient factors that contribute towards patient re-admission to the hospital. Our objective is to construct predictive models that can predict whether a patient is at risk of being re-admitted in near future. In particular, we focus on readmitted patients suffering from pain problems. Pain is a common problem that a patient has to endure even though patient comfort is of utmost importance. A plethora of research has been conducted to lessen pain problems for the patients, though no significant improvements have been made in this regard [22, 23].

To tackle the issue of re-admissions, a lot of research is g. Several techniques and predictive models, which consider several patient factors, including socio-economic status, marital status, sex, and age, among others, have been developed to predict readmission [24, 25]. Some recent research studies have used administrative data to be predict patient readmission within a year [26-28] or even within a month after being discharged from the hospital [24, 29-32]. A few studies have only concentrated on a select group of patients [25, 27, 29, 31] or only on a single hospital [24, 25, 33]. Despite all the work, most of the predictive models have poor predictive capability and are too complex to be utilized in daily practice [26]. Furthermore, none of the aforementioned research studies have considered the effect of nursing care on patient re-admission.

## 2 Data Description

The data for this experiment has been obtained from the HANDS database [34], deployed in 9 units of four different hospitals. The HANDS is an EHR system, designed specifically to record nursing care provided to the patients. Nursing diagnoses were based on NANDA-I [35], outcomes on the Nursing Outcome Classification (NOC) [36] and nursing interventions on the Nursing Interventions Classifications (NIC) [37] terminologies. The data were collected for a period of three years from 2005 till 2008.

In the three year period, there were a total of 42,403 episodes (from 34,927 unique patients).For our analysis purposes, a continuous stay of the patient in a hospital spanning over single or multiple units, is considered as an episode. The episode ends if a patient is discharged or if the patient dies. An episode consists of single or multiple nurse shifts. In our study, we have considered only those episodes with at least two nurse shifts. In every shift, a nurse documents a plan of care (POC). The POC consists of multiple nursing diagnoses (NANDAs), different identified outcomes (NOCs) associated with NANDAs, their initial and expected score (assigned to each unique NOC with value between 1 and 5, 1 being the worst), and interventions (NICs) to achieve the expected outcome. The POC also consists of patients and nurses demographics.

In our dataset, a total of 5298 patients (~15% of the patients in the dataset) were re-admitted, after being discharged, at least once. 2618 of these 5298 patients had either Acute or Chronic Pain (or both) diagnosis in their plan of care (POC) in both the original and re-admission episode. On the other hand, there were 15,956 patients, diagnosed with Pain, that were admitted only once. All patient deaths in hospitals have been excluded in this work. Both the sets were further reduced by considering only patients with NOC: Pain Control. 980 of 2618 patients that were admitted again after being discharged had a NOC of Pain Control, whereas, 5095 of 15956 patients, admitted only one time, had Pain Control as an outcome. Note that the number of single admissions might not be completely accurate, as the patient could have been re-hospitalized to another hospital unit not using HANDS database, or the patient might have been re-admitted after the study period. A few characteristics of the dataset are given in **Error! Reference source not found.**.

**Table 1.** Dataset Characteristics

| Full Dataset Characteristics | |
|---|---:|
| Total number of admissions | 42,203 |
| Number of unique individuals | 34,927 |
| Number of Patients re-admitted | 5298 |
| All patients diagnosed with Pain | 18574 |
| Number of Pain Patients re-admitted | 2618 |
| Number of re-admitted Patients with NOC: Pain Control | 980 |
| Number of single admission patients with NOC: Pain Control | 5095 |
| **Variables used in prediction** | |
| Age (years) mean (SD) | 59.0 (18.4) |
| Length of Stay (hours) mean (SD) | 91.5 (98.5) |
| Average Nurse Experience (years) mean (SD) | 1.7 (2.4) |

## 3 Feature Extraction

In this study, we have conducted predictive modeling at the episode-level. The target variable is whether a patient was re-admitted again or not after being discharged. Hence, the target variable is a binary variable.

The objective of the experiment is to construct predictive models to determine how various factors impact a patient's re-hospitalization. All the predictor variables have been extracted from HANDS dataset and include the following: the initial NOC outcome rating, expected NOC outcome rating at discharge, the actual final NOC outcome rating at discharge (all the ratings for NOC: Pain Control only and vary between 1 and 5), patient's age, length of stay (LOS), NOC: Pain Control met, nurse experience, time of admission and time of discharge. Patients' age, LOS, nurse experience, time of admission and discharge were continuous variables that were discretized for our analysis.

Age has been discretized into four adult groups. These groups, young (18-49), middle-aged (50-64), old (65-84), and very old (85+), are established on theoretical rationale [38] and the data frequency distribution.

We derive LOS for each episode from our database. It was calculated by adding the number of hours for all the nurse shifts in the episode and was grouped into three categories: short (up to 48 hours), medium (48-119 hours), and long stay (120+ hours). Currently, in most places, visits less than 48 hours are called observation visits, considered as short stay. Average LOS of patient in the hospital is typically about 120 hours, and in our study, a stay of between 48 and 120 hours is considered as a medium stay. Anything over 120 hours is, consequently, considered as a long stay [39].

Average nurse experience has also been derived from the database. For nurse experience, a nurse with at least two years of experience in her current position was considered to be an experienced nurse, and nurses with less than two years' experience were considered inexperienced. The episode was categorized as care provided by an experienced nurse team if more than 50% of the nurses providing care in that episode had at least 2 years of experience. These categories were based on professional criteria [40].

NOC being met or not met was calculated using two variables, Expected and Final NOC rating. A NOC is "met" when the final NOC rating is the same or better than the Expected rating, which is set by the nurse that first enters a NOC into a patient's care plan, often when the patient is admitted. Otherwise, NOC is "not met".

The time of Discharge is essentially the completion time of the last POC in the episode and the time is classified into three categories as follows: morning (7am - 3pm), afternoon (3pm – 11pm) and evening (11pm - 7am). These categories were based on timings of nurse shifts in the HANDS database and represent the nursing day, evening and night shifts that are typical in hospitals with 8 hour shifts.

The time of admission is the time when the patient was entered into HANDS database. Like time of discharge, admission time has also been distributed in three different classes or categories: morning (7am - 3pm), afternoon (3pm - 11pm) and evening (11pm - 7am).

Along with these patient and nurse staff variables, the NANDA-I diagnoses and NIC interventions that appeared in the POCs were also considered as predictive variables. The NANDAs and NICs were clustered together by domains and classes, based upon the nursing literature [35, 37]. We included 10 of 12 NANDA-I terms, that had frequencies of more than 5% in our sample episodes. (Activity/Rest, Comfort, Coping/Stress Tolerance, Elimination, Health Promotion, Life Principles, Nutrition, Perception, Role Relationships, and Safety/Protection). Our data sample included terms from all 7 NIC domains (Behavioral, Community, Family, Health System, Safety, Physiological: Basic, and Physiological: Complex). 8 of 19 NANDA classes (Activity/Exercise, Cardiovascular/Pulmonary Responses, Cognition, Hydration, Infection, Physical Comfort, Physical Injury, and Pulmonary System) were included. Finally, of the 30 different NIC classes, 16 (Activity & Exercise Management, Cognitive Therapy, Communication Enhancement, Drug Management, Electrolyte and Acid/Base Management, Immobility Management, Information Management, Nutrition Support, Patient Education, Physical Comfort Promotion, Psychological Comfort Promotion, Respiratory Management, Risk Management, Self-Care Facilitation, Skin/Wound Management, and Tissue Perfusion Management) had frequencies higher than our threshold of 5% in our data sample. A particular NANDA-I or NIC domain and class was assumed to be either present or absent in an episode. In this sparse dataset, the NANDA-I and NIC classes and domains having extremely low frequencies (less than 5%) were excluded to reduce the impacts of spurious correlations.

## 4    Data Modeling

Our key objective for this study was to construct predictive models that can predict whether a patient having pain problems will be re-admitted, after being discharged from a hospital unit. The secondary objective was to assess the feasibility of constructing predictive models using data from a nursing database system.

After extracting and refining the data, multiple models were built on the dataset using different prediction tools and their performances were compared. The models were based on Decision Trees (DT) [41], k-nearest neighbors (k-NN) [42], support-vector machines (SVM) [43], and Naïve-Bayes [44].

## 5    Experimental Results

The analysis was performed to build models for predicting re-hospitalization of patients suffering from Pain problems based on a number of patient and nurse features, nursing diagnoses, and nursing interventions. The performance of the models was evaluated using 10-fold cross-validation [45]. For the experiment, we chose a sample of 2300 patients including both single and multiple admissions. The results are presented in Table 2.

**Table 2.** Model Comparison for experiment predicting re-admission

| Model | Accuracy | Recall | Precision | F Measure | AUC |
|---|---|---|---|---|---|
| Decision Tree | 73.7 | 77.4 | 72.5 | 0.75 | 0.78 |
| Naïve-Bayes | 69.3 | 72.7 | 67.4 | 0.70 | 0.71 |
| K-NN (K = 2) | 64.5 | 84.1 | 60.8 | 0.71 | 0.62 |
| K-NN (K = 5) | 66.1 | 85.1 | 60.5 | 0.71 | 0.69 |
| K-NN (K = 10) | 64.9 | 82.3 | 59.1 | 0.69 | 0.67 |
| SVM | 65.1 | 80.7 | 59.1 | 0.68 | 0.65 |

The preliminary results indicate the Decision Tree and Naïve-Bayes algorithms have a relatively high prediction accuracy compared to k-NN and SVM models. The decision tree has the best accuracy at 73.7%, followed by Naïve-Bayes with an accuracy of 69.3%. K-NN models have accuracy ranging between 64.5-66.1% and SVM has an accuracy of 65.1%. F-measure is 0.75 for the decision tree model and around 0.7 for rest of the models. Area under the curve (AUC) was 0.78 for the decision tree.

Figs. 1-4 give the decision trees after partitioning the tree based on topmost node or the best predictor; age. These figures show the different important variables for these four different groups of patients. **Error! Reference source not found.** shows the decision tree for the young patients. Around 70% of the younger patients had a single admission. The next best predictor is the length of stay. When the LOS is short, around 76.5% of the patients are not re-admitted; for medium LOS, 67.9% have only a single admission; on the other hand, only 57.1% of the young patients with a long episode were not re-admitted.

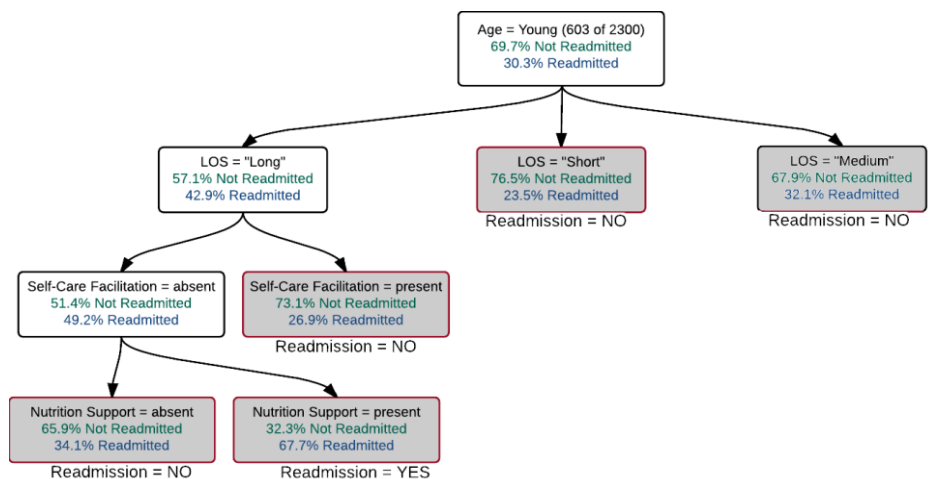

**Fig. 1.** Readmission Decision Tree for Young Patients

The decision tree for middle-aged patients is depicted in **Error! Reference source not found.**. The next best predictor for middle-aged patients is also patient's LOS. When the LOS for middle-aged patient is "short", 69.7% of them are not re-admitted. These set of patients have Final NOC Rating as the next predictor. If the Final Rating is 3 or less, only 43.2% of the patients are not re-admitted. On the other hand, if the Final Rating is 4 or 5, 72.4% of the patients had a single visit only. Whenever the LOS is "medium", 55.3% of the patients are not re-admitted. The next predictor is also the rating in the Final shift for this set. For patients with "long" LOS, 55.7% of the patients were not re-admitted. For these patients, the next predictor was Nutrition NANDA domain. Whenever the Nutrition domain was present, 63.6% of the patients had a single hospital visit, whereas when Nutrition was absent, only 36.9% of the patients were not hospitalized again.

The old patients' decision tree is given in **Error! Reference source not found.**. The number of old patients that had a single admission (50.9%) only and those who were re-admitted (49.1%) was almost equal. When the predictor Behavioral NIC domain was absent, it is observed that 59.3% of the patients were re-admitted as compared to 46.8% when the Behavioral NIC domain was present. For the patients that did not have Behavioral NIC domain, whenever the Expected rating for NOC: Pain Control was 2 or below, all of the patients were re-admitted, though there were only 23 such cases. For patients with behavioral domain present, whenever the NOC: Pain Control was met, around 37.5% of the patients were re-admitted, whereas 54.5% of the patients were re-admitted when the NOC: Pain Control was not met for patients having interventions from the Behavioral domain.

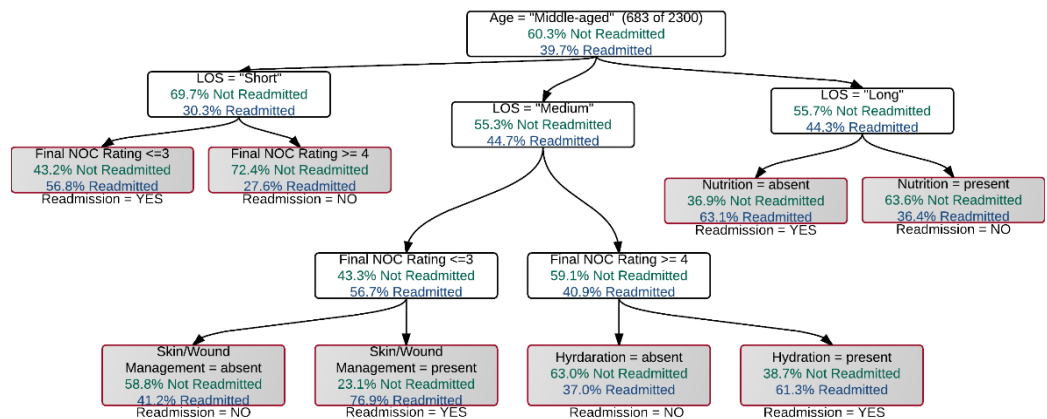

**Fig. 2.** Readmission Decision Tree – Middle-aged Patients

Around 37.9% of the very old patients were not re-admitted (**Error! Reference source not found.**). Whenever a very old patient was discharged during afternoon or evening hours, there was a 66% and 60% chance respectively of the patient being re-admitted again. On the other hand, if the patient was discharged during morning hours, the chance of re-admission was only 35.3%.

The best accuracy results were obtained using the decision tree model. Also, the model generated by the decision tree is easily understandable. Naïve-Bayes model also had a high accuracy, however, the Naïve-Bayes models have independent features assumption [46], and it is often not clear if the features are truly independent. Therefore, we propose to use decision tree predictive modeling on our data.

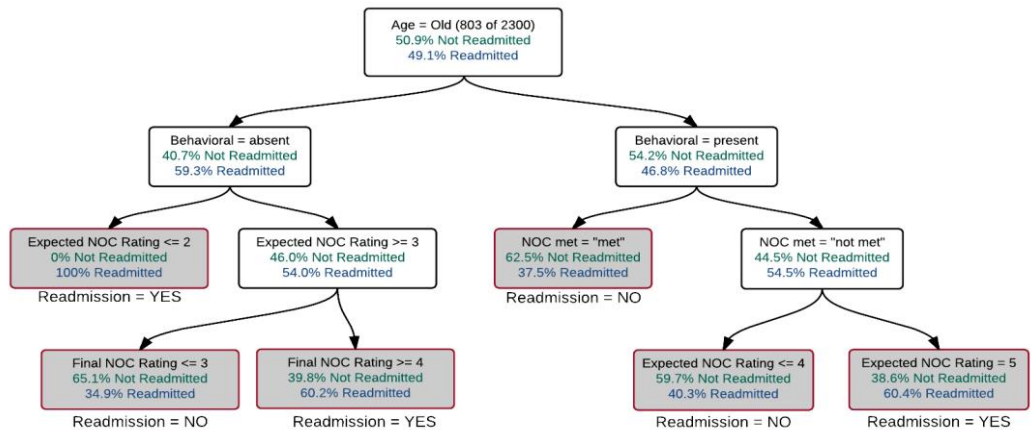

**Fig. 3.** Readmission Decision Tree for Old Patients

## 6      Discussion and Conclusion

Lowering the re-hospitalization rate is one of the main actions that can help achieve a reduction in healthcare costs. The re-admission problem needs to be handled as many hospitals are facing financial issues [47-49]. Different strategies can be implemented using results from predictive modeling. The capability to recognize patients at high risk of re-admission is the key first step to improve quality of care for the patients [50], potentially leading to interventions tailored to individual patients to lower the risk of re-admission. Unfortunately, most of the current work cannot be utilized properly due to different complexities.

In this work, we constructed models to predict hospital re-admission of patients suffering from pain problems using nursing data. Unlike some previous studies, our data were not gathered through questionnaires and interviews, as patients have been known to under-report hospital re-admissions [51]. Decision tree model had the best accuracy of all the models tested and therefore will be used for further analysis. Our preliminary findings suggest that patient demographics, different nursing diagnoses and interventions, among other variables can be used to predict whether a patient will be coming back for treatment. The model had a reasonable accuracy of 73.7%.

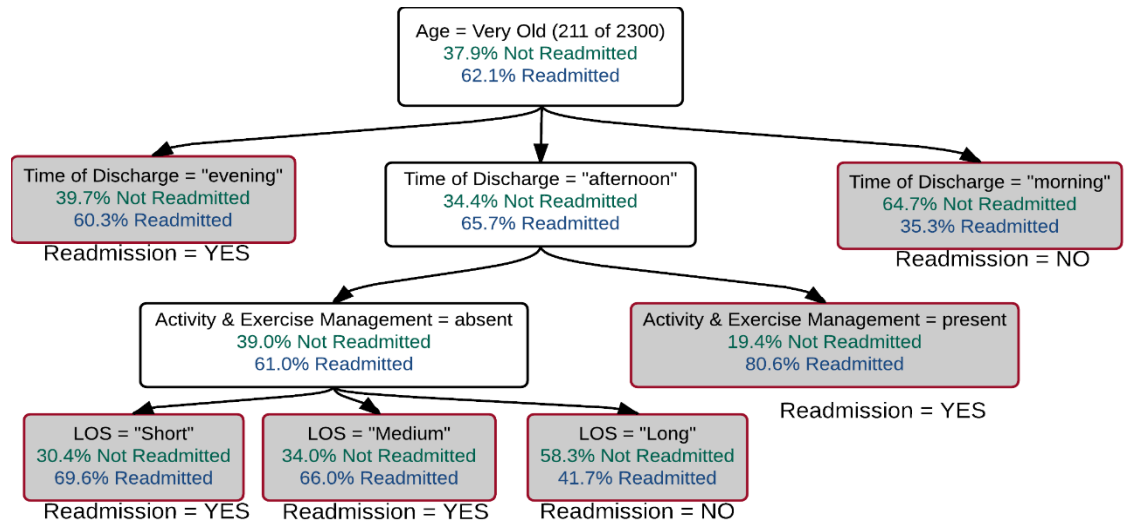

**Fig. 4.** Re-admission Decision Tree for Very Old Patients

The model has some limitations due to data issues. It was developed using data from a nursing EHR system which was not deployed in all units or hospitals. There is a strong probability that some patients that have been counted as a single admission patient, might have been re-admitted to a different hospital unit in which the EHR system was not deployed or was not a part of our original study. Furthermore, a patient might have been re-admitted to a hospital after the study period. Nevertheless, we believe that these differences would not have a considerable effect on the accuracy of the decision tree model.

Notwithstanding a few shortcomings, the predictive modeling techniques have vital implications for developing effective strategies for preventing repeat hospitalization of the patients, since the results of the models can be used to identify at-risk patients of future re-admission the hospital. Further, our use of nursing care data in this analysis has revealed the potential importance of utilizing nursing care variables to identify risk factors of re-admission. This makes sense to us since nurses are the main front line providers of care in the hospital setting. A lot of money can be saved by reducing the re-admission rate at the hospitals, thus careful identification of the risk factors is important. Apart from this benefit, the findings from the current study can be used to improve the care provided to the patients.

# 7 References

1. Hilbert, M. and P. López, *The world's technological capacity to store, communicate, and compute information.* science, 2011. **332**(6025): p. 60-65.
2. Chen, Y., D. Hu, and G. Zhang, *Data Mining and Critical Success Factors in Data Mining Projects*, in *Knowledge Enterprise: Intelligent Strategies in Product Design, Manufacturing, and Management*. 2006, Springer. p. 281-287.
3. Pan, F., et al., *CARPENTER: Finding closed patterns in long biological datasets*, in *International Conference on Knowledge Discovery and Data Mining*. 2003.
4. Liu, H., et al., *Mining Frequent Patterns from Very High Dimensional Data: A Top-Down Row Enumeration Approach*, in *2006 SIAM International Conference on Data Mining (SDM'06)*. 2006: Bethesda, MD. p. 280-291.
5. Jolliffe, I., *Principal component analysis*. 2002: Wiley Online Library.
6. Cox, T.F. and M.A. Cox, *Multidimensional scaling*. 2000: CRC Press.
7. Hyvärinen, A., J. Karhunen, and E. Oja, *Independent component analysis*. Vol. 46. 2004: John Wiley & Sons.
8. Harper, E. and J. Sensmeier. *Why is Big Data Important to Nurses?* Himss 2015 [cited 2015 September 10]; Available from: http://www.himss.org/News/NewsDetail.aspx?ItemNumber=43374.
9. Kavilanz, P.B. *Health care's big money wasters*. 2009 August 10, 2009 [cited 2014 April 29]; Available from: http://www.money.cnn.com/2009/08/10/news/economy/healthcare_money_wasters/.
10. Cuckler, G., *National Health Expenditures Projections 2012-2022*, C.f.M.a.M.S. 2014, Editor. 2014.
11. Smith, P.C., *Performance measurement for health system improvement: experiences, challenges and prospects*. 2009: Cambridge University Press.
12. Billings, J., et al., *Impact of socioeconomic status on hospital use in New York City.* Health affairs, 1993. **12**(1): p. 162-173.
13. Goodman, D.C., et al., *After hospitalization: a Dartmouth atlas report on post-acute care for Medicare beneficiaries.* The Dartmouth Institute, September, 2011. **28**.
14. Yam, C., et al., *Measuring and preventing potentially avoidable hospital readmissions: a review of the literature.* Hong Kong medical

journal= Xianggang yi xue za zhi/Hong Kong Academy of Medicine, 2010. **16**(5): p. 383-389.

15. Herzog, R. *5 Ways Healthcare Providers Can Reduce Costly Hospital Readmissions*. 2013 March 31, 2013 [cited 2015 August 29]; Available from: http://hitconsultant.net/2013/03/31/5-ways-healthcare-providers-can-reduce-costly-hospital-readmissions/.
16. Vest, R.J., et al., *Determinants of preventable readmissions in the United States: A systematic review.* Implementation Science, 2010. **5**(88).
17. *Reducing Hospital Readmission with Enhanced Patient Education*, K.P. Education, Editor. 2010.
18. Jencks, S.F., M.V. Williams, and E.A. Coleman, *Rehospitalizations among patients in the Medicare fee-for-service program.* The New England Journal of Medicine, 2009. **360**: p. 1418-28.
19. Commission, M.P.A., *Report to the Congress: promoting greater efficiency in Medicare*. 2007: Medicare Payment Advisory Commission (MedPAC).
20. Minott, J. *Reducing Hospital Readmissions*. 2008 [cited 2015 August 28]; Available from: http://www.academyhealth.org/files/publications/ReducingHospitalReadmissions.pdf.
21. Foster, D. and G. Harkness. *Healthcare reform: Pending Changes to Reimbursement for 30-day Readmissions.* August 2010 [cited 2015 August 31]; Available from: http://www.communitysolutions.com/assets/2012_Institute_Presentations/acareimbuesementchanges051812.pdf.
22. *A controlled trial to improve care for seriously ill hospitalized patients. The study to understand prognoses and preferences for outcomes and risks of treatments (SUPPORT). The SUPPORT Principal Investigators.* Jama, 1995. **274**(20): p. 1591-8.
23. Yao, Y., et al., *Current state of pain care for hospitalized patients at end of life.* Am J Hosp Palliat Care, 2013. **30**(2): p. 128-36.
24. Hasan, O., et al., *Hospital readmission in general medicine patients: a prediction model.* Journal of general internal medicine, 2010. **25**(3): p. 211-219.
25. Mudge, A.M., et al., *Recurrent readmissions in medical patients: a prospective study.* Journal of Hospital Medicine, 2011. **6**(2): p. 61-67.
26. Billings, J., et al., *Case finding for patients at risk of readmission to hospital: development of algorithm to identify high risk patients.* Bmj, 2006. **333**(7563): p. 327.

27. Cui, Y., et al., *Development and validation of a predictive model for all-cause hospital readmissions in Winnipeg, Canada.* Journal of Health Services Research and Policy, 2015. **20**(2): p. 83-91.
28. Howell, S., et al., *Using routine inpatient data to identify patients at risk of hospital readmission.* BMC Health Services Research, 2009. **9**: p. 96.
29. Holloway, J., S. Medendorp, and J. Bromberg, *Risk factors for early readmission among veterans.* Health services research, 1990. **25**(1 Pt 2): p. 213.
30. Meldon, S.W., et al., *A Brief Risk-stratification Tool to Predict Repeat Emergency Department Visits and Hospitalizationsin Older Patients Discharged from the Emergency Department.* Academic Emergency Medicine, 2003. **10**(3): p. 224-232.
31. Rowland, K., et al., *The discharge of elderly patients from an accident and emergency department: functional changes and risk of readmission.* Age and ageing, 1990. **19**(6): p. 415-418.
32. van Walraven, C., et al., *Derivation and validation of an index to predict early death or unplanned readmission after discharge from hospital to the community.* Canadian Medical Association Journal, 2010. **182**(6): p. 551-557.
33. Phillips, R.S., et al., *Predicting emergency readmissions for patients discharged from the medical service of a teaching hospital.* Journal of general internal medicine, 1987. **2**(6): p. 400-405.
34. Keenan, G., et al., *Maintaining a consistent big picture: Meaningful use of a Web-based POC EHR system.* International Journal of Nursing Knowledge, 2012. **23**(3): p. 119-133.
35. Association, N.A.N.D., *NANDA Nursing Diagnoses*. 2007: North American Nursing Diagnosis Association.
36. Moorhead, S., M. Johnson, and M. Maas. *Iowa Outcomes Project, Nursing outcomes classification (NOC)*. 2004. St. Louis, MO: Mosby.
37. Bulechek, G.M., H.K. Butcher, and J.M. Dochterman, *Nursing interventions classification (NIC)*. 2008: Mosby.
38. Gronbach, K.W., *The Age Curve: How to Profit from the Coming Demographic Storm*. 2008.
39. *Hospital utilization (in non-federal short-stay hospitals)*. 2014: Centers for Disease Control and Prevention.
40. Benner, P., *From novice to expert.* American Journal of Nursing, 1982. **82**(3): p. 402-407.
41. Quinlan, J., *C4.5: Programs for machine learning*, M. Kaufmann, Editor. 2003: San Francisco, CA.

42. Aha, D., D. Kibler, and M. Albert, *Instance-based learning algorithms.* Machine Learning, 1991. **6**(1): p. 37-66.
43. Cortes, C. and V. Vapnik, *Support-vector networks.* Machine Learning, 1995. **20**(3): p. 273.
44. Pearl, J. *Bayesian networks*. in *The handbook of brain theory and neural networks*. 1998. MIT Press.
45. Witten, I.H. and E. Frank, *Data Mining: Practical Machine Learning Tools and Techniques*. 2 ed, ed. J. Gray. 2005: Elsevier.
46. Lewis, D., *Naive (Bayes) at Forty: The Independence Assumption in Information Retrieval*, in *Proceedings of 10th European Conference on Machine Learning*. 1998. p. 4-15.
47. Whitmarsh, C. *Hospitals Facing Economic Challenges*. [cited 2015 September 6]; Available from: http://www.businesslife.com/articles.php?id=1104.
48. Gugliotta, G., *Rural hospitals, beset by financial problems, struggle to survive*, in *The Washington Post*. 2015.
49. Campbell, D., *NHS cuts: One in three hospitals face financial crisis as result of cash squeeze*, in *The Guardian*. 2013.
50. Desikan, P., et al. *Predictive Modeling in Healthcare: Challenges and Opportunities*. [cited 2014 September 27]; Available from: http://lifesciences.ieee.org/publications/newsletter/november-2013/439-predictive-modeling-in-healthcare-challenges-and-opportunities.
51. Norrish, A., et al., *Validity of self-reported hospital admission in a prospective study.* American journal of epidemiology, 1994. **140**(10): p. 938-942.